\documentclass[12pt]{article}

\usepackage{amssymb}
\usepackage{amsfonts}
\usepackage{amsmath}

\input{tcilatex}

\begin{document}

\title{Quantum Vacuum Heuristics}

\author{Howard E. Brandt \\
U.S. Army Research Laboratory\\ Adelphi, MD\\
hbrandt@arl.army.mil}
\date{January 27, 2003}
\maketitle

\begin{abstract} Didactic heuristic arguments, based on the quantum mechanics of the
vacuum and the structure of spacetime, are reviewed concerning particle creation from the
vacuum by an electric field, vacuum radiation in an accelerated frame, black-hole
radiation, minimum mass black holes, spacetime breakdown, maximal proper acceleration,
the spacetime tangent bundle, and intrinsic Planck-scale regularization
of quantum fields.
\end{abstract}

\section{\protect\bigskip INTRODUCTION}

During a break at the\textit{\ First Feynman Festival }[1], Marlan Scully cornered me
with a technical question concerning the Bogoliubov transformation connecting modes of a
quantum field in the derivation of the Hawking radiation from a black hole. In the course
of our discussions, I offered him some heuristic arguments concerning Hawking radiation,
and which I had found to be useful in motivating related ideas based on the quantum
mechanics of the vacuum and the structure of spacetime [2-20]. To my considerable
satisfaction, Marlan then invited me to talk about this at the
\textit{33rd Winter Colloquium on the Physics of Quantum Electronics} [21]. The present
paper documents that talk.

\section{PAIR PRODUCTION FROM THE VACUUM}

In 1950, Schwinger showed that the probability per unit time and per unit volume of
electron-positron pair production from the vacuum by an external electric field $E$ is
given by [22]
\begin{equation}
\frac{d^{4}P}{dtd^{3}r}=\left( \frac{e^{2}E^{2}}{\pi ^{2}\hbar ^{2}c}
\right) \sum\limits_{n=1}^{\infty }\frac{e^{-nE_{c}/E}}{n^{2}},
\end{equation}
where $e$ is the charge of the electron, $\hbar $ is Planck's constant
divided by $2\pi$, $c$ is the speed of light, $dtd^{3}r$ is an infinitesimal volume
element of spacetime, and the characteristic field $E_{c}$, for which copious pair
production occurs, is
\begin{equation}
 E_{c}=\frac{\pi m^{2}c^{3}}{e\hbar },
\end{equation}
where $m\ $is the the mass of the electron.

The characteristic field $E_{c}$ can also be estimated by simple heuristic arguments
[4,23] involving the quantum mechanics of the vacuum. By the time-energy uncertainty
principle, virtual electron-positron pairs of mass $m$ occur in the vacuum fluctuation
during a time

\begin{equation}
\Delta t\ \sim \ \frac{\hbar }{\Delta E}\ \sim \
\frac{\hbar }{2mc^{2}},
\end{equation}
where $\Delta E$ is the energy uncertainty occurring during the time
$\Delta t$. The limiting velocity of light allows this to occur over a distance
\begin{equation}
\Delta x\sim c\Delta t\sim \frac{\hbar }{2mc},
\end{equation}
of the order of the Compton wavelength of the electron. If during the
time $\Delta t$, an external electric field $E$ acts with force $eE$ on each of the
virtual particles over the distance $\Delta x$, doing work
\begin{equation}
\left( eE\right) \left( \hbar /2mc\right) \ \sim \ mc^{2},
\end{equation}
thereby depositing energy $mc^{2}$, then the particle can become real.
Solving Eq.\thinspace (5) for the electric field, one obtains the approximate
characteristic field,
\begin{equation} E\ \sim \ \frac{2m^{2}c^{3}}{e\hbar }\sim \ E_{c}=\frac{\pi
m^{2}c^{3}}{e\hbar },
\end{equation}
to be compared with Eq.\thinspace (2).

\section{UNRUH RADIATION}

If instead of the electric force acting on the vacuum, one analogously considers the
inertial force $ma$ acting on each virtual particle of mass $m$ in the vacuum in a frame
with proper acceleration $a$, one has, analogous to the argument in Section 2 [4,24],
\begin{equation}
\left( ma \right) \left( \hbar /2mc \right) \sim E \sim mc^{2},
\end{equation} 
where $E$ is now the energy of the created particle. (In Minkowski
spacetime, the proper acceleration of a particle is the acceleration in its instantaneous
rest frame.) Solving Eq.\thinspace (7) for $a$, one obtains
\begin{equation} 
a\ \sim \ \frac{2mc^{3}}{\hbar }
\end{equation} 
for the characteristic proper acceleration at which copious production of
particles of mass $m$ from the vacuum will occur. Furthermore, if one takes the energy of
the created particles to be thermally distributed with temperature $T$, as in a
relativistic gas, then Eq.\thinspace (7) becomes
\begin{equation}
\left( ma\right) \left( \hbar /2mc\right) \sim \ 3kT.
\end{equation}
Solving Eq.\thinspace (9) for $T$, one obtains
\begin{equation} T\ \sim \ \frac{\hbar a}{6kc}\ \sim \ T_{u}=\frac{\hbar
a}{2\pi kc},
\end{equation} 
which approximates the Davies-Unruh temperature $T_{u}$ of the vacuum in
an accelerated frame with proper acceleration $a$ [25-27]. The radiated particles
constitute the Unruh radiation from the vacuum.

For the case of massless particles of frequency $\omega /2\pi$ and energy $\hbar \omega
$, Eq.\thinspace (3) is replaced by
\begin{equation}
\Delta t\ \sim \ \frac{\hbar }{\Delta E}\ \sim \
\frac{\hbar }{ 2\hbar \omega }\ \sim \ \frac{1}{2\omega }.
\end{equation}
It then follows that Eq.\thinspace (4) is replaced by
\begin{equation}
\Delta x\ \sim \ c\Delta t\ \sim \ \frac{c}{2\omega }\
\sim \
\frac{\lambda }{4\pi },
\end{equation} 
where $\lambda $ is the wavelength of each particle. Since the effective
relativistic inertial mass of each particle is $\hbar \omega /c^{2}$, and the distance
over which the inertial force can act is given by Eq.\thinspace (12), it follows that,
for massless particles, Eq.\thinspace (9) is to be replaced by [4,24]
\begin{equation}
\left[ \left( \frac{\hbar \omega }{c^{2}}\right) a\right] \left(
\frac{ \lambda }{4\pi }\right) \ \sim \ 3kT.
\end{equation}
Next substituting the massless particle dispersion relation,
\begin{equation}
\omega =\frac{2\pi c}{\lambda },
\end{equation} 
in Eq.\thinspace (14), one again obtains
\begin{equation} 
T\ \sim \ \frac{\hbar a}{6kc}\ ,
\end{equation} 
as in Eq.\thinspace (10).

\section{HAWKING RADIATION}

Next consider a Schwarzschild black hole [2]. The radius of the event horizon of the
black hole is given by its Schwarzschild radius,
\begin{equation} 
R=\frac{2GM}{c^{2}}\bigskip ,
\end{equation} 
where $M$ is the the mass of the black hole, and $G$ is the universal
gravitational constant. The acceleration due to gravity at the event horizon is well
known to be
\begin{equation} 
a_{R}=\frac{1}{\sqrt{g_{00}}}\left( \frac{GM}{R^{2}}\right) ,
\end{equation} 
where $g_{00}$ is the time-time component of the Schwarzschild metric
tensor. Next invoking the equivalence principle, and using Eqs.\thinspace (10), (16), and
(17), one obtains
\begin{equation} 
T_{R}\sim \frac{\hbar a_{R}}{6kc}\sim
\frac{1}{\sqrt{g_{00}}}
\left( \frac{\hbar c^{3}}{24kGM}\right)
\end{equation}%
for the radiation temperature at the event horizon. To obtain the
temperature\ $T_{\infty }$ measured by a distant observer (at infinity), one next invokes
the Tolman relation [28] which requires that $\sqrt{g_{00}}\,T$ \ be constant at every
point in an equilibrium spacetime for proper temperature $T$, and since $g_{00}=1$ at
infinity, one obtains
\begin{equation} 
T_{\infty }=\sqrt{g_{00}}\,T_{R}.
\end{equation} 
Then substituting Eq.\thinspace (18) in Eq.\thinspace (19), one obtains
[2]
\begin{equation}
T_{\infty }\sim \frac{\hbar c^{3}}{24kGM}\sim T_{H}=\frac{
\hbar c^{3}}{8\pi kGM},
\end{equation} 
the temperature $T_{H}$ of the black hole due to its Hawking radiation
[29]. [The exact equality for $T_{H}$ also follows from the equality in Eq.\thinspace
(10), together with Eqs.\thinspace (17) and (19) [2].]
 
Another process results from tidal forces, and also yields the same
temperature of the black hole. This might have been anticipated long ago
because Schr\"{o}dinger had argued, already in 1939, that spacetime
curvature (which produces tidal forces) can cause particle production
[30]. For the Schwarzschild black hole, the work done by the tidal forces
acting on a particle pair, for a time allowed by the time-energy
uncertainty principle, and during which pair creation can occur, can be
approximated by
\begin{eqnarray}
\left[ \frac{1}{\sqrt{g_{rr}}}\frac{d}{dr}\left( \frac{- GMm}{\sqrt{g_{00}}
r^{2}}\right) \left( \sqrt{g_{rr}}\frac{\hbar }{2mc}\right)
\right] \left(
\frac{1}{2}\sqrt{g_{rr}}\frac{\hbar }{2mc}\right)	\nonumber	\\
\sim
\frac{1}{g_{00} }\left[ \left( \frac{GMm}{r^{3}}\right) \left( \frac{\hbar
}{2mc}\right) 
\right] \left( \frac{\hbar }{2mc}\right) ,
\end{eqnarray}
since for the Schwarzschild metric $g_{rr}=1/g_{00}$. Evaluating
Eq.\thinspace (21) at the Schwarzschild radius, Eq.\thinspace (16), and equating this
work to the rest energy
$2mc^{2}$ of the created pair, one has
\begin{equation}
\frac{1}{g_{00}}\left\{ \left[ \frac{GMm}{\left\{
\frac{2GM}{c^{2}}\right\} ^{3}}\right] \left( \frac{\hbar }{2mc}\right) \right\} \left(
\frac{
\hbar }{2mc}\right) \sim \ 2mc^{2}.
\end{equation} 
Solving Eq.\thinspace (22) for the particle rest energy $mc^{2}$, and
taking it again to be thermal, one obtains
\begin{equation}
\frac{1}{\sqrt{g_{00}}}\left( \frac{\hbar c^{3}}{8GM}\right) \
\sim \ mc^{2}\ \sim \ 3kT_{R}.
\end{equation}
Therefore, the temperature at the event horizon is
\begin{equation} T_{R}\ \sim \ \ \frac{1}{\sqrt{g_{00}}}\left( \frac{\hbar
c^{3}}{24kGM }\right) ,
\end{equation}
and again invoking the Tolman relation, Eq.\thinspace (19), one again obtains
Eq.\thinspace (20).

\section{SPACETIME BREAKDOWN}

One might ask, on the basis of Eq.\thinspace (8), for what proper acceleration $a$ would
copious production of black hole pairs from the vacuum occur [2,4,31,32]. If the created
particles are to be black holes, then their Schwarzschild radius, Eq.\thinspace (16),
must exceed their extent (Compton wavelength), given by Eq.\thinspace (4), and thus one
has
\begin{equation}
\frac{2Gm}{c^{2}}\geq \frac{\hbar }{2mc},
\end{equation}
and therefore
\begin{equation} m\geq \frac{1}{2}\left( \frac{\hbar c}{G}\right) ^{1/2}.
\end{equation}
The minimum value of the mass for which the created particle is a black
hole is therefore
\begin{equation} m_{0}\sim \frac{1}{2}\left( \frac{\hbar c}{G}\right) ^{1/2},
\end{equation}
which is of the order of the Planck mass. Thus letting the mass in
Eq.\thinspace (8) be $m_{0}$, one obtains
\begin{equation}
a_{0}\ \sim \ \frac{2m_{0}c^{3}}{\hbar }.
\end{equation}
In a frame with proper acceleration given by Eq.\thinspace (28), copious
production of black holes from the vacuum\ would occur, resulting in breakdown of the
classical spacetime structure due to the drastic alteration of its topology [4,31]. [As
argued in Section 7, each created black hole of mass $m_{0}$ corresponds to one unit of
topology per Planck volume.]

\section{MAXIMAL PROPER ACCELERATION}

Substituting Eq.\thinspace (27) in Eq.\thinspace (28), one obtains [2]
\begin{equation}
a_{0}\ =\ 2\pi \alpha \left( \frac{c^{7}}{\hbar G}\right) ^{1/2},
\end{equation}
where $\alpha $ is a number of order unity that can be shown to follow
from string theory [5]. It is compelling to define $a_{0}$, given by Eq.\thinspace (29),
as the universal maximum possible proper acceleration relative to the vacuum because, for
that value of the proper acceleration, spacetime would break down due to copious black
hole production, and the very concept of acceleration would have no meaning because of
the resulting complex topological structure.

At maximal proper acceleration, Eq.\thinspace (29), the temperature of the vacuum Unruh
radiation, Eq.\thinspace (10), becomes [2,4,5]
\begin{equation}
T_{\text{max}}=\frac{\hbar a_{0}}{2\pi kc}=\frac{\alpha }{k}\left(
\frac{ \hbar c^{5}}{G}\right) ^{1/2}=T_{\text{S}},
\end{equation}
where $T_{\text{S}}$ is the Sakharov temperature, the maximum possible
temperature of anything [2,4,33]. In 1966 Sakharov, using heuristic arguments involving
the equation of state of black-body radiation at the extreme density of one quantum per
Planck volume, concluded that this temperature (of the order of the Planck temperature)
is the maximum possible temperature of thermal radiation, or anything else in equilibrium
with it [33].

It can also be shown that at maximal proper acceleration $a_{0},$ the string vacuum
becomes unstable, the functional integral expressing the response of a particle detector
becomes singular, and any attempt to increase the acceleration above
$a_{0}$ simply results in string lengthening, with the proper acceleration remaining at
$a_{0}$, and the temperature remaining at the Sakharov temperature
$T_{\text{S }}$[5,34-37]. The maximal proper acceleration, Eq.\thinspace (29) was first
obtained in 1982 by the author by directly equating the Davies-Unruh temperature, the
temperature of the vacuum in an accelerated frame, to the Sakharov temperature, the
maximum possible temperature of thermal radiation [2].

\section{MINIMUM-MASS BLACK HOLE}

Equations (20) and (30) also suggest a minimum possible mass of a black hole, namely that
of a black hole at the Sakharov temperature [2],
\begin{equation}
M_{0}=\frac{1}{8\pi \alpha }\left( \frac{\hbar c}{G}\right) ^{1/2},
\end{equation}
which is of of the order of the Planck mass, consistent with
Eq.\thinspace (27). The corresponding minimum radius of the event horizon of a black hole
follows from Eqs.\thinspace (16) and (31), namely,
\begin{equation}
R_{0}=\frac{2GM_{0}}{c^{2}}=\frac{1}{4\pi \alpha }\left(
\frac{\hbar G}{ c^{3}}\right) ^{1/2},
\end{equation}
which is of the order of the Planck length. Thus, each minimum-mass black
hole corresponds to one unit of topology [38] per Planck volume [2]. Also, for the minimum
entropy of a black hole (with the area of its event horizon expressed in Planck units),
\begin{equation}
S_{0}=\frac{1}{4}\left( \frac{4\pi R_{0}^{2}}{\hbar G/c^{3}}\right) k,
\end{equation}
one obtains, by substituting Eq.\thinspace (32) in Eq.\thinspace (33),
the value,
\begin{equation}
S_{0}=\frac{1}{4}\left( \frac{1}{4\pi \alpha ^{2}}\right) k=\left(
\frac{1}{16\pi \alpha ^{2}}\right) k,
\end{equation}
which is of the order of Boltzmann's constant.

\section{MINIMUM RADIUS OF CURVATURE OF WORLDLINES}

According to the differential geometry of spacetime, the proper acceleration, $a$, of a
particle on a world line in curved spacetime is given by [4]
\begin{equation} a^{2}=-c^{4}g_{\mu \nu }\frac{Dv^{\mu }}{ds}\frac{Dv^{\nu }}{ds}.
\end{equation}
In Eq.\thinspace (35), $g_{\mu \nu }$ is the metric tensor of spacetime,
the four-velocity $v^{\mu }$ of the particle is given by
\begin{equation}
\ v^{\mu }=\frac{dx^{\mu }}{ds},
\end{equation}
and $\frac{Dv^{\mu }}{ds}$ denotes the covariant derivative of the
four-velocity with respect to the interval along the particle's worldline in spacetime,
namely,
\begin{equation}
\frac{Dv^{\mu }}{ds}=\frac{dv^{\mu }}{ds}+\Gamma _{\alpha \beta }^{\mu }v^{\alpha
}v^{\beta },
\end{equation}
where $\Gamma _{\alpha \beta }^{\mu }$ is the affine connection, and
\begin{equation}
ds^{2}=g_{\mu \nu }dx^{\mu }dx^{\nu },
\end{equation}
is the line element of spacetime. The radius of curvature of the
worldline is
\begin{equation}
\rho =\frac{c^{2}}{a}.
\end{equation}
It then follows from Eqs.\thinspace (29) and (39) that since $a_{0}$ is
the maximum possible value of $a$, the minimum radius of curvature of world lines is
given by [39]
\begin{equation}
\rho _{0}=\frac{c^{2}}{a_{0}} = \frac{1}{2\pi \alpha }\left(
\frac{\hbar G}{c^{3}}\right) ^{1/2},
\end{equation}
which is of the order of the Planck length.

In earlier work, the author has explored some of the differential geometric implications
of the limiting acceleration, Eq.$\,$(29) and the minimum radius of curvature of
worldlines, Eq.\thinspace (40). This has included extensive analyses of the differential
geometric structure of the tangent bundle of spacetime [3-20].

\section{SPACETIME TANGENT BUNDLE}

It is argued in Section 6 that $a_{0}$, given by Eq.\thinspace (29), is the universal
maximum possible proper acceleration of anything relative to the vacuum. Hence, for any
proper acceleration $a$, one requires
\begin{equation}
a^{2}\leq a_{0}^{2}.
\end{equation}
Then substituting Eqs.\thinspace (35) and (37) in Eq.\thinspace (41), one
obtains
\begin{equation}
-c^{4}g_{\mu \nu }\left( \frac{dv^{\mu }}{ds}+\Gamma _{\alpha
\beta }^{\mu }v^{\alpha }v^{\beta }\right) \left( \frac{dv^{\nu }}{ds}+\Gamma _{\lambda
\delta }^{\nu }v^{\lambda }v^{\delta }\right) \leq a_{0}^{2}.
\end{equation}
Next substituting Eqs.\thinspace (36), (40), and (38) in Eq.\thinspace
(42), one obtains [4,40]
\begin{equation}
d\sigma ^{2}\equiv g_{\mu \nu }dx^{\mu }dx^{\nu }+\rho _{0}^{2}g_{\mu \nu
}\left( dv^{\mu }+\Gamma _{\alpha \beta }^{\mu }v^{\alpha }dx^{\beta }\right) \left(
dv^{\nu }+\Gamma _{\lambda \delta }^{\nu }v^{\lambda }dx^{\delta }\right) \geq 0.
\end{equation}
Equation (43) defines the eight-dimensional quadratic form $d\sigma^{2}$, which is
nonnegative along a worldline. The inequality, Eq.\thinspace (43), simply expresses the
fact that the proper acceleration of any object can never exceed the maximal proper
acceleration. By analogy with the construction of the spacetime line element of general
relativity from the limiting speed of light, it is natural to take $d\sigma ^{2}$ to be
the line element in the tangent bundle of spacetime, in which the spacetime coordinates
$x^{\mu }$ are the coordinates in the spacetime base manifold, and the four-velocity
coordinates $\rho _{0}v^{\mu }$ (modulo a factor of $\rho _{0}$) are the tangent space
coordinates.

The bundle line element, Eq.\thinspace (43) can be rewritten as follows
[4-7,9,40]:
\begin{equation} d\sigma ^{2}=G_{MN}dx^{M}dx^{N},\ \ \ \ \ \{M,N=1,2,...8\},
\end{equation}
where the bundle coordinates are
\begin{equation}
\left\{ x^{M}\right\} \equiv \left\{ x^{\mu },\rho _{0}v^{\mu }\right\} ,\ \
\ \ \{M=1,2,...8;\ \mu =0,1,2,3\},
\end{equation}
and the metric of the tangent bundle of spacetime is
\begin{equation}
G_{MN}=\left[
\begin{array}{cc}
g_{\mu \nu }+g_{\alpha \beta }A_{\ \mu }^{\alpha }A_{\ \nu }^{\beta } &
A_{n\mu } \\ A_{m\nu } & g_{mn}
\end{array}
\right] ,
\end{equation}
where
\begin{equation} A_{\ \nu }^{\mu }=\rho _{0}v^{\lambda }\Gamma _{\ \lambda \nu }^{\mu }.
\end{equation}
The bundle metric $G_{MN}$, given by Eq.\thinspace (46), has a structure
similar to that of an eight-dimensional Kaluza-Klein gauge theory in which the higher
dimensions are in four-velocity space, and $A_{\ \nu }^{\mu }$ is the gauge potential.
Equations (44)-(47) served as the starting point for investigating possible implications
of a limiting proper acceleration for the differential geometric structure of the tangent
bundle of spacetime [3-20].

\section{PLANCK-SCALE REGULARIZATION OF QUANTUM FIELDS}

One implication of the limiting proper acceleration, which the author recently began to
explore is an intrinsic regularization of quantum fields at the Planck scale [15-20]. In
this work, a scalar quantum field is defined on the spacetime tangent bundle with
vanishing eigenvalue when acted on by the invariant Laplace-Beltrami operator defined on
the bundle. For the case of a Minkowski spacetime in the base manifold of the spacetime
tangent bundle, it follows that the scalar quantum field is given by [15-20]
\begin{eqnarray}
\phi (x,v)=2\int \frac{d^{3}\mathbf{p}}{(2\pi )^{3/2}\left( 2p^{0}N\right) ^{1/2}}\left[
e^{-ipx/\hbar }e^{-\rho _{0}pv/\hbar }\theta
\left( \rho _{0}pv/\hbar \right) a(p)	\nonumber\right.\\
+ \left. e^{ipx/\hbar }e^{\rho _{0}pv/\hbar }\theta
\left( -\rho _{0}pv/\hbar \right) a^{\dag }(p)\right] ,
\end{eqnarray}
where $p$ denotes the familiar particle four-momentum $p^{\mu}=\{p^{0},
\mathbf{p\}}$; $\mathit{x}$ denotes the spacetime coordinates
$x^{\mu}=\{x^{0},\mathbf{x\}}$; $v$ denotes the four-velocity coordinates
$v^{\mu}=\{v^{0},\mathbf{v\}}$; $\rho _{0}$ is given by Eq.\thinspace (40);
$\theta$ is the Heaviside step function; $N$ is an appropriate normalization factor that
approaches unity in the mathematical limit of infinite maximal proper acceleration;$\
$and $a^{\dag }(p)$ and $a(p)$ are the familiar particle creation and annihilation
operators. In the mathematical limit of infinite maximal proper acceleration, the scalar
field, Eq.\thinspace (48), has the same form as a canonical Lorentz-invariant scalar
quantum field (as it must). It can be shown that both the positive and negative frequency
terms in Eq.\thinspace (48) are proportional to [15-20]
\begin{equation}
e^{-\rho _{0}\left\vert pv\right\vert /\hbar }=\exp \left[ -
\frac{1}{2\pi
\alpha }\frac{\gamma m}{m_{pl}}\left( \left[ 1+\left(
\frac{\left\vert
\mathbf{p}\right\vert }{mc}\right) ^{2}\right] ^{1/2}-
\frac{\mathbf{p}\cdot d
\mathbf{x/}dt}{mc^{2}}\right) \right] ,
\end{equation}
where $m$ is the particle mass;
\begin{equation}
m_{pl}=\left( \frac{\hbar c}{G}\right) ^{1/2}
\end{equation}
is the Planck mass; $d\mathbf{x/}dt$ denotes the velocity coordinates for
the quantum field in the tangent space; and
\begin{equation}
\gamma =\left[ 1-\frac{\left\vert d\mathbf{x/}dt\right\vert ^{2}}{c^{2}}
\right] ^{-1/2}.
\end{equation}
Thus the quantum field is intrinsically regularized at the Planck scale,
and elementary particle excitations exceeding the Planck energy are exponentially
suppressed. This suppression is evidently not inconsistent with the arguments of Sections
5 and 6.

\section{ACKNOWLEDGEMENTS}

The author wishes to thank Marlan Scully for inviting him to present this paper at the
\textit{33rd Winter Colloquium on the Physics of Quantum Electronics}, held 5-9 January
2003 in Snowbird, Utah. This work was supported in part by the U.S. Army Research
Laboratory.

\section{REFERENCES}

\ \ \ \ \thinspace \lbrack 1] \textit{First Feynman Festival}, 23-27 August 2002, Physics
Department, University of Maryland, College Park, MD.

[2] H. E. Brandt, Lett. Nuovo Cim. \textbf{38}, 522 (1983); \textbf{39}, 192 (1984).

[3] H. E. Brandt, Nucl. Phys. B (Proc. Suppl.) \textbf{6}, 367 (1989).

[4] H. E. Brandt, Found. Phys. Lett. \textbf{2}, 39, 405 (1989).

[5] H. E. Brandt, Found. Phys. Lett. \textbf{4}, 523 (1991).

[6] H. E. Brandt, Found. Phys. \textbf{21}, 1285 (1991).

[7] H. E. Brandt, Found. Phys. Lett. \textbf{5}, 43 (1992).

[8] H. E. Brandt, Found. Phys. Lett. \textbf{5}, 221 (1992).

[9] H. E. Brandt, Int. J. Theor. Phys. \textbf{31}, 575 (1992).

[10] H. E. Brandt, Found. Phys. Lett. \textbf{5}, 315 (1992).

[11] H. E. Brandt, Found. Phys. Lett. \textbf{6}, 245 (1993).

[12] H. E. Brandt, Found. Phys. Lett. \textbf{6}, 339 (1993).

[13] H. E. Brandt, Found. Phys. Lett. \textbf{7}, 297 (1994).

[14] H. E. Brandt, Contemp. Math. \textbf{196}, 273 (1996).

[15] H. E. Brandt, Found. Phys. Lett. \textbf{11}, 265 (1998).

[16] H. E. Brandt, Chaos, Solitons \& Fractals \textbf{10}, 267 (1999).

[17] H. E. Brandt, Found. Phys. Lett. \textbf{13}, 307 (2000).

[18] H. E. Brandt, Found. Phys. Lett. \textbf{13}, 581, (2000).

[19] Rep. Math. Phys. \textbf{45}, 389 (2000).

[20] H. E. Brandt, ``Finslerian Fields," in \textit{Finslerian Geometries: A Meeting of
Minds}, P. L. Antonelli, ed. (Kluwer, Dordrecht, 2000) pp. 131-138.

[21] \textit{33rd Winter Colloquium on the Physics of Quantum Electronics}, 5-9 January
2003, Snowbird, Utah.

[22] J. Schwinger, Phys. Rev. \textbf{82}, 664 (1951).

[23] V. L. Ginzburg, \textit{Key Problems of Physics and Astrophysics}, MIR Publishers,
Moscow (1978), p.77.

[24] H. E. Brandt, Bull. Am. Phys. Soc. \textbf{27}, 538 (1982); \textbf{28}, 722 (1983).

[25] P. C. W. Davies, J. Phys. A \textbf{8}, 609 (1975).

[26] W. G. Unruh, Phys. Rev. D \textbf{14}, 870 (1976).

[27] N. D. Birrell and P. W. Davies, \textit{Quantum Fields in Curved Space,} Cambridge
University Press, Cambridge, 1984.

[28] R. C. Tolman, \textit{Relativity, Thermodynamics and Cosmology, Dover, NY (1987),
pp. 312-319.}

[29] S. W. Hawking, Nature (London) \textbf{24}, 30 (1974); Commun. Math. Phys.
\textbf{43}, 199 (1975).

[30] E. Schr\"{o}dinger, Physica (Utrecht) \textbf{6}, 899 (1939).

[31] H. E. Brandt, Bull. Am. Phys. Soc. \textbf{30}, 341 (1985).

[32] H. E. Brandt, "Maximal-Acceleration Invariant Phase Space," in \textit{The Physics
of \ Phase Space}, Y. S. Kim and W. W. Zachary, eds. (Springer-Verlag, Berlin, 1987) pp.
414-416.

[33] A. D. Sakharov, JETP Lett. \textbf{3}, 388 (1966).

[34] N. Sakai, ``Hawking Radiation in String Theories," in \textit{Particles and Nuclei},
H. Terazawa, ed. (World Scientific, Singapore, 1986) pp. 286-296.

[35] R. Parentini and R. Potting, Phys. Rev. Lett. \textbf{63}, 945 (1989).

[36] M. J. Bowick and S. B. Giddings, Nucl. Phys. B \textbf{325}, 631 (1989).

[37] S. B. Giddings, Phys. Lett. B \textbf{226}, 55 (1989).

[38] S. W. Hawking, Nucl. Phys. B \textbf{144}, 349 (1978).

[39] H. E. Brandt, Bull. Am. Phys. Soc. \textbf{29}, 651 (1984).

[40] H. E. Brandt, "The Maximal Acceleration Group," in \textit{XIIIth International
Colloquium on Group Theoretical Methods in Physics}, W. W. Zachary, ed. (World
Scientific, Singapore, 1984) pp. 519-522.

\end{document}